\documentclass[12pt]{article}
\usepackage[dvips]{graphicx}
\begin{document}
\newcommand{\x}{\vec{r}}
\newcommand{\mc}[1]{\multicolumn{2}{c}{#1}}
\newcommand{\mcd}[1]{\multicolumn{2}{c||}{#1}}

\title{Fast and accurate molecular Hartree-Fock with a finite-element multigrid method}
\author{O.Beck$^1$\footnote{becko@physik.uni-kassel.de}, D.Heinemann$^2$, D.Kolb$^1$\footnote{kolb@physik.uni-kassel.de}\\$^1$ Fachbereich Naturwissenschaft, Universit\"at Kassel, 34132 Kassel, Germany\\$^2$ Verwaltungs-DV, Universit\"at Kassel, 34132 Kassel, Germany}
\maketitle
\begin{abstract}
We present a multigrid scheme for the solution of finite-element Hartree-Fock equations for diatomic molecules. It is shown to be fast and accurate, the time effort depending linearly on the number of variables. Results are given for the molecules LiH, BH, N$_2$ and for the Be atom in our molecular grid which agrees very well with accurate values from an atomic code. Highest accuracies were obtained by applying an extrapolation scheme; we compare with other numerical methods. For N$_2$ we get an accuracy below 1 nHartree.
\end{abstract}
\section{Introduction}
For benchmark calculations in Hartree-Fock it is necessary to have a very good accuracy. This may be achieved through numerical methods like finite differences \cite{laak,pyyk87,sund,pyyk89,pyyk91,kobus93,kobus94,moncrieff95,kobus96,moncrieff98,kobus99,kobus00,kobus01}. Recently, the accuracy has been pushed to the sub-$\mu$Hartree range by employing more than 10000 points in a 2-dimensional finite difference grid\cite{kobus01}. Alternatively a finite-element program was developed which gave quite accurate results \cite{hei88,hei90} with less points. However, the time effort of this method scales quadratically in the number of unknowns. 

In recent papers by Kopylow et al.\cite{kopA,kopB} it was shown that a multigrid approach for solving finite element equations applied to the Kohn-Sham equations of density functional theory gives fast and accurate results. 

Extrapolation methods \cite{brez,walz,stoer} are a tool to gain better accuracy from a sequence of values for a given property for which the asymptotic behaviour is known. They have been successfully applied to MP2 correlation energies for closed shell atoms \cite{flores} where several orders of magnitude have been gained. More recent is the application to one-electron Dirac-FEM solutions for diatomics \cite{kullie} which yields also several orders gain in accuracy.

In this paper we present a scheme which utilizes multigrid techniques \cite{hack,bran} and scales linear in the number of variables. Extrapolation methods are employed in addition in order to further improve the accuracy.

In part \ref{sec:hf} we describe the Hartree-Fock approximation used. In part \ref{sec:coor} our coordinate transformation is given. In part \ref{sec:fem} we state the discretization of the Hartree-Fock equations by the finite-element method. And in part \ref{sec:mg} we explain the multigrid scheme that is used. In part \ref{sec:res} we give results both directly calculated and extrapolated for Be, BH, LiH and N$_2$ and compare with other work\cite{laak,kobus93,kobus01}.
\section{Hartree-Fock Method}
\label{sec:hf}
Microscopic physical systems like molecules are described by wavefunctions whose behaviour is governed by the Hamiltonian of the system. Often one is only interested in the energy levels of the system which are given through the eigenvalue equation
\begin{equation} \label{schr}
H\Psi = E \Psi
\end{equation}
The eigenfunction $\Psi$ is a many-body wavefunction. Equation \ref{schr} can be solved analytically for a few physical systems only. For others it is necessary to make approximations like the Hartree-Fock method where a single determinant serves as the ansatz function for the many-body wave function(Slater determinant).
For electronic systems the Hamiltonian is
\begin{equation}
H = \sum_{i=1}^N T_i + \sum_{i=1}^N V_i + \sum_{i<j} V_{ee}(|\x_i-\x_j|)
\end{equation}
where $T_i$ denotes the kinetic energy operator $-\frac{\hbar^2}{2m} \nabla_i^2$ of a single electron, $V_i = V_{ext}(\x_i)$ is a given external potential and $V_{ee}$ the interaction potential between electrons.
The variation of the single particle wave functions $\phi_j$(orbitals) in the Slater determinant results in the Hartree-Fock equations:
\begin{eqnarray} \label{hf}
\left\{ \hat{T} + V_{ext}(\x) + V_{Dir}(\x) \right\} \phi_j (\x) - \sum_i V_{ji}(\x) \phi_i (\x) = \epsilon_j \phi_j (\x) \\ \label{v_ex}
\Delta V_{ji} (\x) = -4\pi \rho_{ji}(\x)\\ \label{v_dir}
\Delta V_{Dir} (\x) = -4\pi \rho (\x)
\end{eqnarray}
where $\rho (\x) = \rho (\x,\x)$ is the density, the diagonal part of the density matrix $\rho (\x,\x') = \sum_i \phi_i^*(\x') \phi_i(\x)$ and $\rho_{ji}(\x)= \phi_j(\x) \phi_i^*(\x)$ the exchange density.
We now define
\begin{math} \label{Vx}
g_x^j = \sum_i V_{ji} (\x) \phi_i (\x)
\end{math}
and $H_{lok}$ = $\hat{T} + V_{ext}(\x) + V_{Dir}(\x)$.
Equation \ref{hf} then takes the form
\begin{equation} \label{hf-eq}
H_{lok} \phi_j(\x) - g_x^j = \epsilon_j \phi_j(\x)
\end{equation} 
Equations \ref{v_ex}-\ref{hf-eq} are solved iteratively (SCF iteration) in the following way. First we take a trial set of functions {$\phi_i$}, compute $V_{Dir}$ and $V_{ji}$ and get $H_{lok}$ and $g_x^j$. Then one calculates an (approximate) eigenvalue $\tilde{\epsilon}_j$:
\begin{equation} \label{eig}
\tilde{\epsilon}_j = \int \phi_j^* (H_{lok} \phi_j - g_x^j) d^3r
\end{equation}
and solves directly the inhomogeneous equation system
\begin{equation}  \label{hf-eq1}
(H_{lok} -\tilde{\epsilon}_j ) \phi_j = g_x^j ,j=1,\dots,N
\end{equation}
in order to get new wave functions \{$\phi_i$\}. Then these new wave functions are orthogonalized from the functions with the lowest to the highest eigenvalues $\tilde{\epsilon}_i$ of occupied states(Gram Schmidt method). For these wavefunctions the total energy is computed as the expectation value for the corresponding Slater determinant. From these \{$\phi_j$\} new $V_{Dir}$, $V_{ji}$ and $H_{lok}$, $g_x^j$ are computed and equations \ref{eig}, \ref{hf-eq1} solved again. This process is re-iterated till a wanted accuracy in the total energy or the energy levels of the orbitals is obtained. In order to get stable and fast convergence the direct potential is mixed with the old one:
\begin{math}
V_{Dir} = p_{mix} * V_{Dir}^{old} + (1-p_{mix}) * V_{Dir}^{new}
;p_{mix} = 0.95 
\end{math}.

\section{Diatomic molecules}
\label{sec:coor}
Diatomic molecules have axial symmetry, $L_z$ commutes with H and thus leads to good quantum numbers $m_j=l_{z,j}$. We take a restricted Hartree-Fock approach and demand the symmetry of the single-particle wavefunction to have axial symmetry also. This leads to the ansatz in cylindrical coordinates for the orbitals. 
\begin{equation}
\phi_j(r,\varphi,z) = f_j(r,z) e^{im_j\varphi}
\end{equation}
The two-center point nucleus Coulomb singularities are best described by elliptic hyperbolic coordinates:
\begin{eqnarray}
\xi  & = & \frac{r_1+r_2}{R}\\
\eta & = & \frac{r_1-r_2}{R}
\end{eqnarray}
where $r_k$ is the distance to th $k$-th nucleus(k=1,2) and R the internuclear distance. These coordinates remove the singularity of the Coulomb potential which is necessary for a high order convergence behaviour of the finite-element method. For the computation of best energies in closed shell systems the prolate spheroidal coordinates are favorably used which emerge after a singular coordinate transformation(the back transform is singular at the nuclear centers, i.e. at $\xi$=1(s=0), $\eta = \pm 1 (t=0,\pi)$).
\begin{eqnarray}
\xi  & = & \cosh s\\
\eta & = & \cos t 
\end{eqnarray}
This singular transform improves the analytic properties of polynomial ansatz functions in s,t considerably in a finite-element approach even though it leads to a problematic behaviour at the inner boundaries $\xi$=1 (s=0) and $\eta$=$\pm$1(t=0,$\pi$) respectively. In FEM this difficulty is adequately handled by an open boundary treatment. On the outer boundary we use a closed boundary with boundary values 0.

\section{Finite element method}
\label{sec:fem}
In the finite element method a variational formulation is generally the starting point. In our case, the variational integral corresponding to equation \ref{hf-eq1} is:

\begin{equation}
I = \int  \phi_j^* (\frac{1}{2}(H_{lok} - \tilde{\epsilon}_j) \phi_j - g_x^j) d^3r 
\end{equation}

In the finite-element method(FEM) the space is subdivided into several subspaces called elements on which locally defined formfunctions $N_i$ are used as approximation. Via the FEM ansatz $\phi(\x) = \sum_i c_i N_i(\x)$ and variation with respect to the $c_i$ the following equation results: 

\begin{eqnarray}
\sum_l (H_{lok,kl} & - & \tilde{\epsilon}_j S_{kl}) c_l^{(j)} = g_{x,k}^{(j)}\\
H_{lok,kl} & = & \int N_k^*(\x) H_{lok} N_l(\x) d^3r\\
g_{x,k}^{(j)} & = & \int N_k^*(\x) g_x^j d^3r\\
S_{kl} & = & \int N_k^*(\x) N_l(\x) d^3r\\
\end{eqnarray}

At the outer boundary the wavefunctions are set to zero(closed boundary), at the inner boundaries(symmetry axis) the wavefunctions may take any values(open boundary). 
Analogously there exists a variational integral for equation \ref{v_ex}:

\begin{equation}
I = \int \left[ -\frac{1}{2}|\nabla V_{ij}|^2 + 4\pi \rho_{ij}V_{ij}^* \right] d^3r
\end{equation}
A corresponding integral exists for the direct potential from equation \ref{v_dir} which then can be treated in the same way.
This leads after insertion of the FEM ansatz to the linear inhomogeneous equation systems

\begin{eqnarray}
 \sum_k D_{lk} V_k^{ij}  = \rho_l^{ij} \\
\mbox{with }  D_{lk} =  \int \nabla N_l^*(\x) \nabla N_k(\x) d^3r \\
\mbox{and }  \rho_l^{ij} =  4\pi \int \rho_{ij}(\x) N_l^*(\x) d^3r
\end{eqnarray}

The outer boundary values for the potentials $V_{ij}$ are computed from a multipole expansion of the densities $\rho_{ij}$ up to the fifth order. The ansatz functions $N_k$ are complete polynomials of order p (p $\le$ 7) in s,t and thus complicated transcendental functions of the real space vector $\x$; p is called the order of the elements and is normally taken to be 4. At the element boundaries continuity at the grid points is demanded which due to the serendip properties of polynomials leads to continuity over the whole element boundaries.
It can be shown that for large numbers of points n the error due to the finite number of points is proportional to $\frac{1}{n^p}$\cite{femconv}.
\section{Multigrid method}
\label{sec:mg}
\begin{figure}[ht]
\caption{multigrid algorithm} \label{mg}
\begin{center}
F: finest grid; C: coarsest grid
\hspace*{-1cm}
\includegraphics[height=18.0cm,angle=-90]{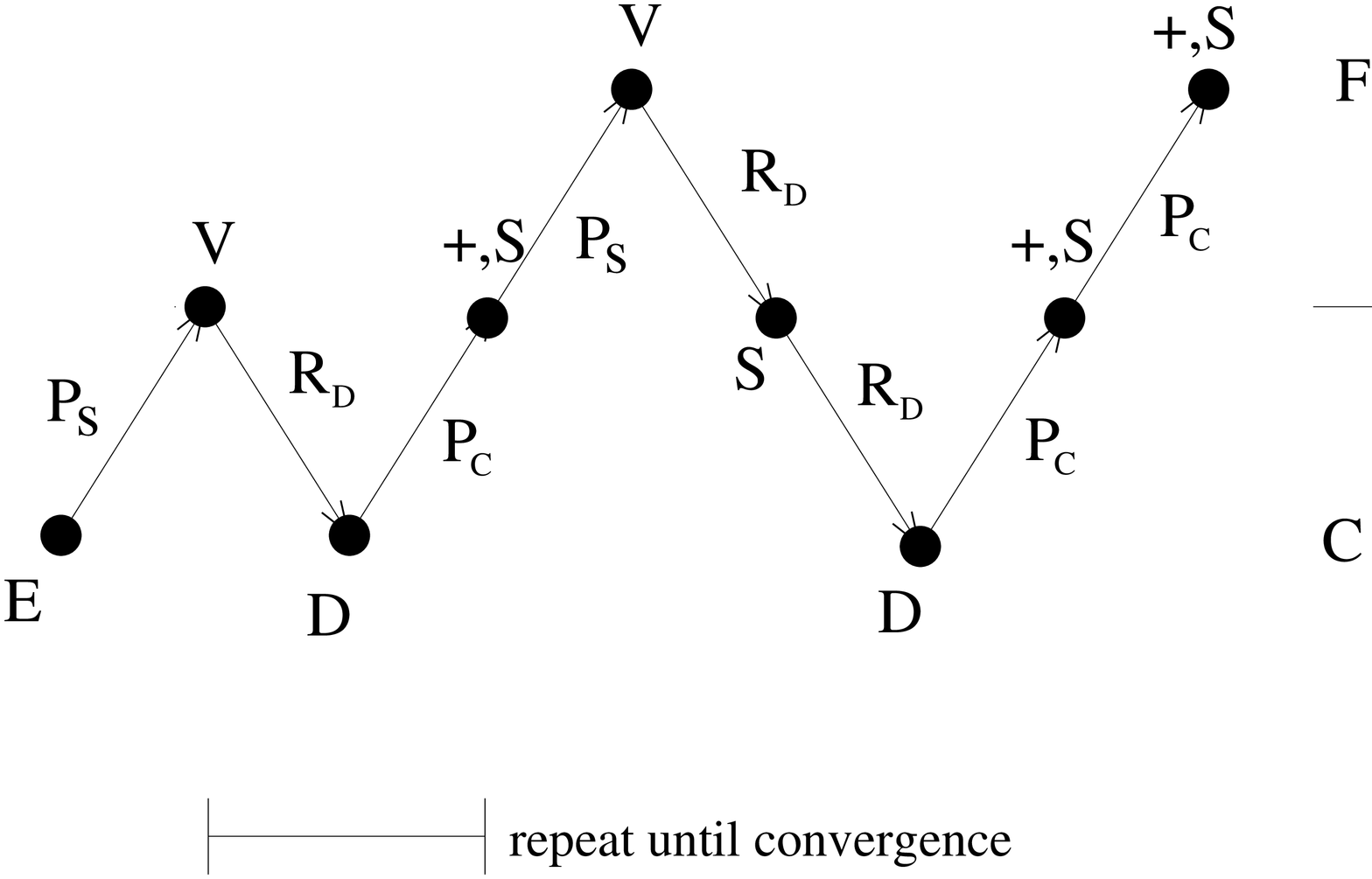}
\end{center}
\end{figure}

In order to solve the FEM matrix equations we use a multigrid scheme. The multigrid method \cite{hack,bran} combines the good smoothing behaviour of iterative methods(e.g. Gauss-Seidel, CG) with an effective elimination of long-range errors which are treated at coarser grids. 
In diagram \ref{mg} we show the MG scheme we use. First we solve directly on the coarsest grid(E). Then this solution is prolongated to the next finer grid($P_S$). Prolongation is an interpolation step to the finer grid values where we use all ansatz functions over each coarse element. If we have a solution from an older scf cycle we compare the defects of both and take the one with the smaller one(V). If this vector is not converged we do a V-cycle: First we restrict the defect to the next smaller grid($R_D$). The restriction step is done by the transpose of the interpolation matrix $P_S$,$R_D = P_S^T$ and is thus of the same (high) order as the interpolation. If this is not the coarsest grid, we smooth and restrict the resulting defect to the next coarser grid(S,$R_D$). This is repeated till the coarsest grid is reached where we solve directly for the defect(D). The solution of the defect is prolongated to the next finer grid and added to the old approximation vector($P_C$,+). This new vector is then smoothed(S). This is repeated till the level from which the V-cycle started is reached. If the vector is not converged well enough we repeat the V-cycles till sufficient convergence. Then we prolongate the solution vector($\mbox{P}_S$) to the next finer grid and repeat the whole procedure.
In our computations we used 15 conjugate gradient steps in each smoothing. On the average we needed 2 V-cycles on the finest grid per scf iteration for every state $\phi_j$ and potential $V_{ij}$.

\section{Results and Discussion}
\label{sec:res}
We first present values for BH, LiH and Be using different grids ranging from 545 to 16641 points. These results were extrapolated using both inverse power and rational geometric extrapolation which are described in Flores and Kolb\cite{flores}. For large numbers of points n the energy obeys the following formula:
\begin{equation}
E(n) = E_{\infty} + \frac{C}{n^p},
\end{equation}
where E(n) is the energy for a given number of points, $E_{\infty}$ the exact value, p the polynomial order and C a constant. The same relationship holds for all properties except the densities at the nuclei. For the extrapolation of the densities at the nuclei we took p = 3 for the leading term of the asymptotic expansion. Fitting of three subsequent densities to the formula $ D(n) = D_\infty + \frac{C}{n^p}$ gave a convergence of the parameter p close to 3.

In table \ref{BH} the properties of BH(R=2.336) are given for different numbers of points and the extrapolated values together with the finite difference results of Laaksonen et al.\cite{laak} and, for the total energy and electron levels, Kobus\cite{kobus93}. Compared with those of Laaksonen et al. our directly computed results are generally more accurate up to three digits. The only exception are the densities at the nuclei. For the hydrogen nucleus we are as accurate as Laaksonen et al. and at the heavier nuclei our directly calculated results are about two digits less accurate, however by extrapolation we gain 4 digits. This reflects the fact that the finite element method optimizes integral properties not point-like ones. In most cases our values are different from Laaksonen et al.'s though mostly only in the last digit given by them.

Our results for 16641 points are as accurate as those of Kobus, who had more than three times the number of points(52441). And our extrapolated results should be clearly more accurate.

For LiH(R=3.015) the extrapolated values are given in table \ref{lih+be}. Our results agree with the more recent numbers of Kobus but give 2-3 more digits.
Be is computed with the two-center grid where the Be atom is placed in one center, the other center, R apart, is empty(dummy center). We can compare with the results of the atomic GRASP code\cite{surz}. The total energy is given for 390 points, the energy levels, due to convergence problems, for 220 points. For the total energy our result disagrees slightly from the Kobus values but agrees with the GRASP values which shows our values to be more accurate. For the energy levels the GRASP results agree with both but has to few digits to differentiate between them.

\begin{figure}[p]
\caption{time of solution routines for 1 scf iteration} \label{time}
\center{N$_2$\\solid: for potentials; dashed: for wavefunctions}
\includegraphics[height=12.0cm]{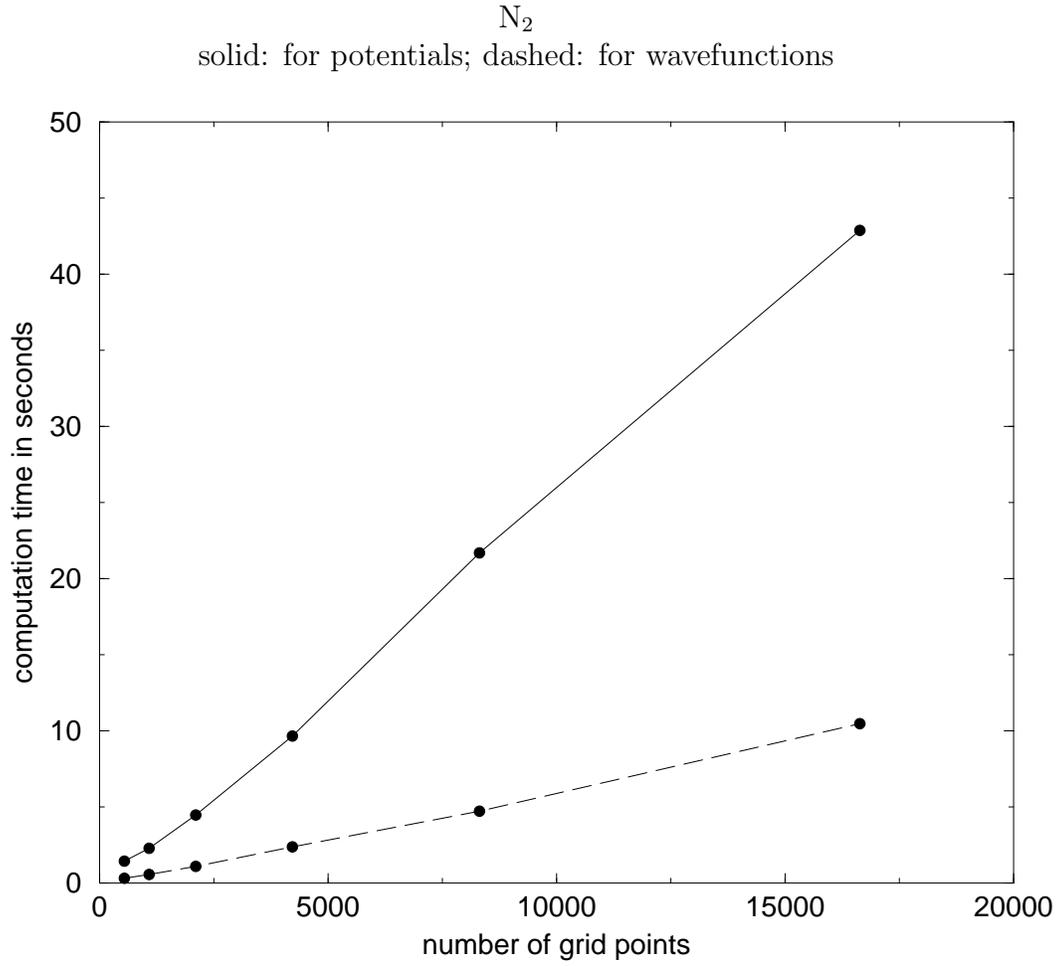}
\end{figure}

In figure \ref{time} the computation times for the solution routines for different numbers of grid points are given. It shows a linear dependence on the number of points. The same is true for the total time per scf iteration. This allows us to use rather large numbers of grid points.
 
In table \ref{n2-p4} the convergence of the total energy of N$_2$(R = 2.068) with respect to the number of points is given up to 148225 points. Here we have the expected convergence behaviour of the finite element method where the leading error term for a high number of grid points is proportional to $\frac{1}{n^p}$(p being the order of the polynomials used). The extrapolated values were computed for one sequence without and one with the 148225 point value. Both are identical apart from the self-consistency error. 
In order to test the accuracy of the results we took the truncation parameter d=18 a.u. instead of 25 a.u. in order to test whether the results were dependent on the properties on the outer boundary. d is the distance between the point on the outermost ellipse $\xi_{max}$ = const. and a focal point, if the distance is taken perpendicular to the symmetry axis. The values show a slightly faster convergence because of the higher density of points in the inner region but converge to the same result for the higher number of points.

\begin{figure}[p]
\caption{convergence for grids with different geometry} \label{ext-diff-geo}
\center{black and white circles denote different geometries\\$p_{eff}$ determined from $\Delta$E = $\frac{C}{n^p}$;in brackets: $p_{eff}$ for same geometry}
\hspace*{-1cm}
\includegraphics[height=12.0cm,angle=0]{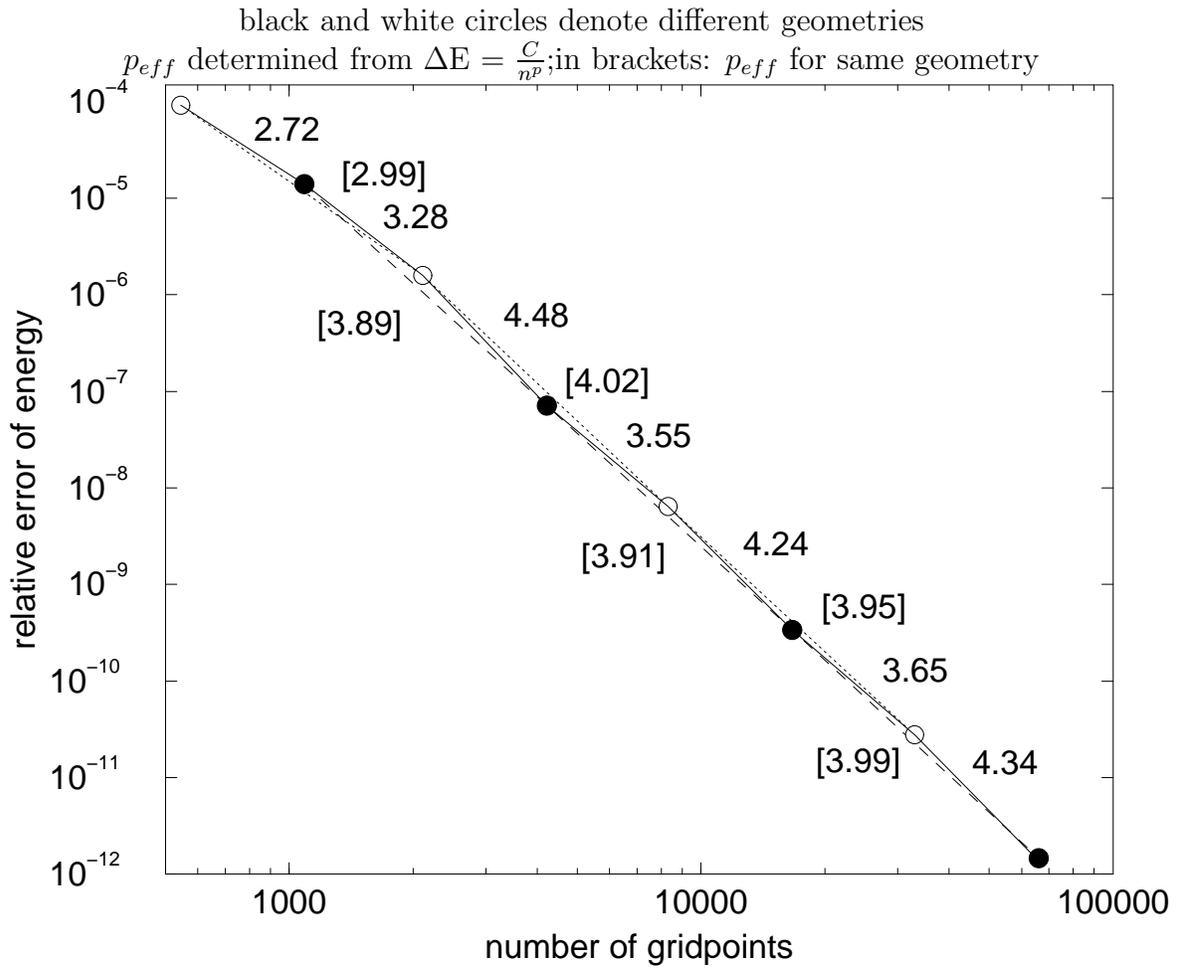}
\end{figure}

In table \ref{n2-pvar} the results for 6th-order polynomials with  d = 40 a.u. and 7th-order polynomials with d = 18 a.u. are shown. No effect from boundary values or from the different orders can be seen up to 1nHartree. The difference of the various orders is probably due to the bigger truncation error accumulation for the higher orders.
This shows our results for the total energy of N$_2$ to be accurate to at least 1 nHartree, and presumably up to two more digits. The finite difference result of Kobus et al. \cite{kobus01} has an error of 13 nHartree for his 793 $\times$ 793 grid lying well in the sub-$\mu$Hartree level of accuracy as was claimed. This accuracy can be reached with our standard method with only 37249 points. For higher orders p it takes only 9409(p=6) or 3613(p=7) points.  
In comparison to the finite difference scheme of Kobus et al.\cite{kobus01}
 we can achieve an accuracy which is by al least 2 orders of magnitude better with much smaller numbers of grid points and thus remarkably small computational times for our high precision benchmark results.
The same holds true with respect to the old finite-element results of Heinemann et al. \cite{hei90}, where we gain 5-7 digits(see table \ref{n2-res}).
It should be noted that our computation with the highest number of points took less than a day on an ordinary personal computer.
At last, we want to point out that extrapolation schemes have to be applied judiciously. In figure \ref{ext-diff-geo} the relative errors of the total energies are given for different grids. Unlike the expected convergence $\propto \frac{1}{n^p}$(p=4) one gets an alternating order parameter p. Closer inspection showed that the grids have alternatingly a different geometry. If grids with the same geometry are taken one gets $p \to 4$ and correspondingly good extrapolation values.
\begin{table}[p]
\caption{Results for BH(R=2.336): total energy, energy levels $\epsilon_i$, multipole moments of order L and densities at the nuclei; all in a.u.} \label{BH}
\hspace*{-2.5cm}
\begin{tabular}{c|r@{.}lr@{.}lr@{.}lr@{.}l}
points & \mc{total energy} & \mc{$\epsilon_1$} & \mc{$\epsilon_2$} & \mc{$\epsilon_3$}\\
\hline
625    &\bf-25&\bf13\rm098274805960 &\bf-7&\bf686\rm16024068801 &\bf-0&\bf6481\rm92901104539 &\bf -0&\bf3484\rm18481279812 \\
1089   &\bf-25&\bf131\rm42119823359 &\bf-7&\bf6862\rm2122038152 &\bf-0&\bf648187\rm699192955 &\bf -0&\bf348423\rm192223713 \\
2401   &\bf-25&\bf1315\rm9271731769 &\bf-7&\bf68626\rm555785907 &\bf-0&\bf6481872\rm93511312 &\bf -0&\bf3484237\rm51988741 \\
4225   &\bf-25&\bf13159\rm798871873 &\bf-7&\bf686267\rm18236676 &\bf-0&\bf64818726\rm8672215 &\bf -0&\bf34842377\rm7833311 \\
9409   &\bf-25&\bf1315986\rm7258947 &\bf-7&\bf6862673\rm6927568 &\bf-0&\bf648187265\rm330006 &\bf -0&\bf348423781\rm443339 \\
16641  &\bf-25&\bf13159869\rm928299 &\bf-7&\bf68626737\rm702293 &\bf-0&\bf6481872652\rm04238 &\bf -0&\bf3484237815\rm82957 \\
\hline
extrapol &\bf-25&\bf13159870231    &\bf -7&\bf68626737794    &\bf -0&\bf6481872651901   &\bf  -0&\bf3484237815982   \\
\hline
Kobus\cite{kobus93} &\bf -25&\bf13159870   &\bf -7&\bf686267370  &\bf -0&\bf648187256       &\bf  -0&\bf348423779       \\
Laak.\cite{laak}    &\bf -25&\bf131647     &\bf -7&\bf686283     &\bf -0&\bf648190          &\bf  -0&\bf348426          \\
\end{tabular}
\hspace*{-2.5cm}
\begin{tabular}{c|r@{.}lr@{.}lr@{.}lr@{.}l}
points & \mc{L=1} & \mc{2}  & \mc{3} & \mc{4} \\
\hline
625      &\bf -5&\bf352\rm6511173 &\bf 12&\bf186\rm3258126 &\bf -15&\bf640\rm9316245 &\bf 24&\bf84\rm81986257 \\
1089     &\bf -5&\bf3524\rm735227 &\bf 12&\bf1862\rm527191 &\bf -15&\bf6411\rm035350 &\bf 24&\bf8492\rm335950 \\
2401     &\bf -5&\bf35246\rm79377 &\bf 12&\bf18624\rm05731 &\bf -15&\bf641093\rm4947 &\bf 24&\bf84925\rm89014 \\
4225     &\bf -5&\bf3524669\rm842 &\bf 12&\bf186240\rm0311 &\bf -15&\bf64109335\rm54 &\bf 24&\bf849260\rm9410 \\
9409     &\bf -5&\bf35246695\rm79 &\bf 12&\bf18623996\rm66 &\bf -15&\bf641093354\rm8 &\bf 24&\bf8492612\rm346 \\
16641    &\bf -5&\bf352466956\rm9 &\bf 12&\bf186239964\rm0 &\bf -15&\bf6410933546  &\bf 24&\bf84926124\rm46 \\
\hline
extrapol &\bf -5&\bf3524669568 &\bf 12&\bf1862399637 &\bf -15&\bf6410933546 &\bf 24&\bf8492612457 \\
\hline
Laak.\cite{laak}     &\bf -5&\bf352466     &\bf 12&\bf18621      &\bf -15&\bf64103      &\bf 24&\bf84888 \\
\end{tabular}
\begin{tabular}{c|r@{.}lr@{.}l}
points & \mc{$\rho(\x_B)$} & \mc{$\rho(\x_H)$} \\
\hline
625      &\bf 6\rm9&179642371 &\bf 0&\bf46\rm683681664 \\
1089     &\bf 7\rm0&644543782 &\bf 0&\bf467\rm41415376 \\
2401     &\bf 71& 536579057      &\bf 0&\bf4675\rm4727884 \\
4225     &\bf 71&\bf6\rm60901149   &\bf 0&\bf46755\rm872297 \\
9409     &\bf 71&\bf69\rm1078340   &\bf 0&\bf467561\rm04352 \\
16641    &\bf 71&\bf69\rm3793197   &\bf 0&\bf4675612\rm2952 \\
\hline
extrapol &\bf 71&\bf694413   &\bf 0&\bf4675612700 \\
\hline
Laak.\cite{laak} &\bf 71&\bf69451     &\bf 0&\bf467561 \\
\end{tabular}
\end{table}

\newpage
\begin{table}[p]
\caption{results for LiH,Be: total energy, energy levels $\epsilon_i$, multipole moments of order L and densities at the nuclei; all in a.u.} \label{lih+be}
\begin{tabular}{c|ccc}
LiH           & extrapolated values&   Laaksonen et al. & Kobus \\
R=3.015 & & & \\
\hline
energy        &   -7.987352237228   & -7.987354 & -7.987352237 \\
$\epsilon_1$  &   -2.445233713306   & -2.445234 & -2.4452337133 \\
$\epsilon_2$  &   -0.301738270249   & -0.301738 & -0.3017382702 \\
L=      1     &   -0.65318943587    & -0.653190 & - \\
        2     &    7.12821973712    &  7.128219 & - \\
        3     &   -2.90955527116    & -2.909556 & - \\
        4     &   16.275742582      & 16.02756  & - \\
$\rho(\x_{Li})$ &   13.789722803      & 13.789729 & - \\
$\rho(\x_H)$    &    0.37406093101    &  0.374061 & - \\
\end{tabular}
\end{table}

\begin{table}[p]
\begin{tabular}{c|cccc}
Be          & extrapolated values&   Laaksonen et al. & Kobus & GRASP \\
            &      R=2.00        &                    & R=2.00&       \\
\hline
energy      & -14.573023168305  & -14.5730226 & -14.573023170 & -14.573023168\\
$\epsilon_1$&  -4.732669897448  &  -4.7326689 & -4.732669898  & -4.7326699 \\
$\epsilon_2$&  -0.3092695515724 &  -0.30926957& -0.3092695522 &  0.30926955 \\
\end{tabular}
\end{table}

\begin{table}[p]
\caption{Total energy of N$_2$(R=2.068)} \label{n2-p4}
\begin{tabular}{c|r@{.}lr@{.}l}
points & \mc{total energy(d=25)} & \mc{total energy(d=18)} \\
\hline
625    &\bf -108&\bf98\rm8278969512   &\bf -108&\bf99\rm0118887918  \\
1089   &\bf -108&\bf99\rm2300729719 5 &\bf -108&\bf99\rm2661731070  \\
2401   &\bf -108&\bf993\rm744526847  &\bf -108&\bf993\rm775395838  \\
4225   &\bf -108&\bf9938\rm1783520   &\bf -108&\bf9938\rm2035051  \\
9409   &\bf -108&\bf993825\rm276408  &\bf -108&\bf993825\rm404018 \\
16641  &\bf -108&\bf9938255\rm97923 5 &\bf -108&\bf9938256\rm10993  \\
37249  &\bf -108&\bf99382563\rm334   &\bf -108&\bf99382563\rm387 \\
66049  &\bf -108&\bf993825634\rm67   &\bf -108&\bf993825634\rm72 \\
\hline
extrapol&\bf -108&\bf99382563482      &\bf -108&\bf99382563482 \\
\hline
148225  &\bf -108&\bf99382563481      &\bf -108&\bf99382563482 \\
\hline
extrapol&\bf -108&\bf99382563482      &\bf -108&\bf99382563482  \\
\hline
Kobus et al.&\bf -108&\bf993825622 \\
\end{tabular}

\end{table}

\begin{table}[hb]
\caption{Total energy of N$_2$ for different orders and d}
\begin{tabular}{c|r@{.}l||c|c} \label{n2-pvar}

points & \mcd{total energy (p=6,d=40)} & points & total energy (p=7,d=18))\\
\hline
1369 & -108&9938095840 & 1849 & -108.9938248739\\
2401 & -108&9938243325 & 3613 & -108.9938256311\\
5329 & -108&9938256092 & 7225 & -108.9938256350\\
9409 & -108&9938256343 &&\\
21025& -108&99382563487 &&\\
37249& -108&993825634866&&\\
\end{tabular}

\end{table}

\begin{table}[hb]
\caption{results for N$_2$: total energy, energy levels $\epsilon_i$, multipole moments of order L and densities at the nuclei; all in a.u.}
\center
\begin{tabular}{c|r@{.}lr@{.}l} \label{n2-res}

points & \mc{extrapolated values} & \mc{Heinemann et al.\cite{hei90}}\\
\hline
energy         & -108&99382563482 & -108&993826 \\
$\epsilon_1$   & -15&68186695242  &  -15&681867 \\
$\epsilon_2$   & -15&67825164397  &  -15&678252 \\
$\epsilon_3$   & -1&473422499578  &  -1&473423  \\
$\epsilon_4$   & -0&778076815628  &  -0&778077  \\
$\epsilon_5$   & -0&6347931345534 &  -0&634793  \\
$\epsilon_6$   & -0&6156250666967 &  -0&615625  \\
L=      2      &  15&908084537079 &  \mc{ } \\
        4      &  23&3942874333   &  \mc{} \\
$\rho(\x_{N})$ &  205&3983861     &  \mc{}\\

\end{tabular}

\end{table}

\section{Acknowledgement}
One of us(O.B.) acknowledges financial support of the Deutsche Forschungsgemeinschaft(DFG).

\end{document}